\documentclass{PoS}
\pdfoutput=1
\usepackage[utf8]{inputenc}
\usepackage[T1]{fontenc}
\usepackage{amsmath,amssymb,slashed}
\usepackage{import}
\usepackage{pgfplots}
\usepackage{textcomp}
\newcommand{\gino}{\lambda}

\newcommand{\aetap}{\text{a--}\eta'}
\newcommand{\api}{\text{a--}\pi}
\newcommand{\afn}{\text{a--}f_0}

\newcommand*{\myfont}{\fontfamily{Palatino}\selectfont}
\DeclareTextFontCommand{\textmyfont}{\myfont}

\newcommand{\fdoi}[2]{\href{http://dx.doi.org/\detokenize{#1}}{#2}}

\title{Continuum limit of SU(3) $\mathcal{N}=1$ supersymmetric Yang-Mills theory and
	supersymmetric gauge theories on the lattice}

\ShortTitle{SU(3) SYM and SUSY gauge theories on the lattice}

\author{Sajid Ali \\
	Institut f\"ur Theoretische Physik, Universit\"at M\"unster, Wilhelm-Klemm-Str.~9, D-48149 M\"unster\\
	Department of Physics, Government College University Lahore, Lahore 54000, Pakistan}
\author{\speaker{Georg Bergner}\\
	University of Jena, Institute for Theoretical Physics\\ 
	Max-Wien-Platz 1, D-07743 Jena, Germany \\
	E-mail: \email{georg.bergner@uni-jena.de}}
\author{Henning Gerber\\
	Institut f\"ur Theoretische Physik, Universit\"at M\"unster\\ 
	Wilhelm-Klemm-Str.~9, D-48149 M\"unster, Germany}
\author{Camilo L\'opez\\
	University of Jena, Institute for Theoretical Physics\\ 
	Max-Wien-Platz 1, D-07743 Jena, Germany}      
\author{Istvan Montvay\\
	Deutsches Elektronen-Synchrotron DESY\\
	Notkestrasse 85, D-22607 Hamburg, Germany}
\author{Gernot M\"unster\\
	Institut f\"ur Theoretische Physik, Universit\"at M\"unster\\ 
	Wilhelm-Klemm-Str.~9, D-48149 M\"unster, Germany}    
\author{Stefano Piemonte\\
	University of Regensburg, Institute for Theoretical Physics\\ 
	Universit\"atstr.~31, D-93040 Regensburg, Germany}
\author{Philipp Scior\\
	Fakult\"at f\"ur Physik, Universit\"at Bielefeld\\
	Universit\"atsstr.~25, D-33615 Bielefeld, Germany}

\abstract{We summarize our investigations of several aspects of $\mathcal{N}=1$ supersymmetric Yang-Mills (SYM) theory. We present our final results for SU(3) $\mathcal{N}=1$ SYM simulated with Wilson fermions. We also discuss the first test of the simulations of the theory with overlap gluinos. Finally, we present some recent progresses concerning the phase structure of the compactified theory on $R^3\times S^1$.}

\FullConference{37th International Symposium on Lattice Field Theory - Lattice2019\\
		16-22 June 2019\\
		Wuhan, China}

\begin{document}
\section{Continuum limit of SU(3) $\mathcal{N}=1$ super Yang-Mills theory with Wilson fermions}
Supersymmetric gauge theories play a central role for several theoretical developments towards an analytical understanding of strong interactions, such as gauge-gravity duality. However, their numerical investigation on the lattice leads to several challenges and unsolved problems. It is the central goal of our investigations to perform numerical simulations of supersymmetric gauge theories. 

The first target of our lattice simulations is the spectrum of bound states of $\mathcal{N}=1$ supersymmetric Yang-Mills theory (SYM), to confirm the consistency with the theoretically expected formation of supermultiplets. In these studies, the theory has been discretized using an efficient and simple approach, applying Wilson fermions. In a first preparatory study we have investigated the theory with gauge group SU(2) \cite{Bergner:2015adz}. More recently we have also completed our studies of the bound state spectrum for SU(3) SYM \cite{Ali:2019agk}.

SYM is the supersymmetric counterpart of Yang-Mills theory and contains the fermionic partners of the gluons, the gluinos ($\gino$). The Euclidan Lagrangian is
\begin{equation}
		\mathcal{L}=\frac{1}{4}
		F_{\mu\nu}F^{\mu\nu}+\frac{1}{2}\bar{\gino}(\slashed{D}+m_g)\gino\; .
\end{equation}
The gluinos are Majorana fermions in the adjoint representation of the gauge group. The Lagrangian $\mathcal{L}$ is invariant under the supersymmetry transformations: $\delta A_\mu=-2\mathrm{i}\, \bar{\lambda} \gamma_\mu \varepsilon$,\linebreak $\delta \lambda=-\sigma_{\mu\nu}F_{\mu\nu}\varepsilon$, in case of a vanishing gluino mass $m_g=0$.

Supersymmetry is broken in any local lattice discretization. Therefore an important aim of our studies is to provide evidence for a restoration of supersymmetry in the continuum limit. We have investigated two signals of supersymmetry: the supersymmetric Ward identities and the formation of mass-degenerate supermultiplets of bound states. The lightest multiplets have been conjectured to be chiral supermultiplets, each consisting of a scalar, a pseudoscalar, and a spin-\textonehalf~fermionic particle. The bosonic particles can be realized by either mesonic states (gluino-balls) or glueballs, while the fermionic partner is a gluino-glue state combining gluino and gluon fields. The proposed mesonic members of a supermultiplet are the $\afn$ ($\gino\gino$) and the $\aetap$ ($\gino\gamma_5\gino$) meson, while the states of glueball type are the $0^{++}$ and $0^{-+}$ glueballs \cite{Veneziano:1982ah,Farrar:1997fn}. In our most recent investigations we have considered in detail the relevant mixing of these two multiplets, and we obtained more reliable results for the lightest states \cite{Ali:2019gzj}. We have also started to investigate other operators and bound states of the theory such as baryonic operators~\cite{Ali:2018bar}. Our recent optimizations of the operator basis have been presented in a separate contribution to this conference.

In the most generic case, supersymmetry can only be obtained by a fine-tuning of the parameters of the lattice action. In the special case of SYM the fine-tuning can be avoided if chiral symmetry is realized on the lattice using Ginsparg-Wilson fermions. In our first investigations we have, however, relied on the simulations with Wilson fermions. These require the fine-tuning of a single parameter, the fermion mass $m_g$ (or equivalently the hopping parameter $\kappa$) \cite{Curci:1986sm}. Our tuning approach relies on the signal for chiral symmetry restoration provided by a vanishing adjoint pion mass. This particle is not a physical state of the theory, but can be defined in a partially quenched setup \cite{Munster:2014cja}. 

Our previous simulations of SU(2) gauge theory have been performed with a Wilson action using stout-smeared gauge links, and tree-level improved gauge action. For our new investigations of SU(3) SYM, we have employed one-loop clover-improvement of the Wilson fermion action and demonstrated that it significantly reduces the lattice artefacts. We started by estimating the best range of simulation parameters from investigations of finite volume effects, the sign problem, and topological freezing that slows down the simulations. The sign problem in SYM with Wilson fermions is due to the Pfaffian of the Dirac operator that results from the functional integration of Majorana fermions. Negative signs are rare in our simulations and can be taken into account by reweighting. In this way, we have estimated a reliable parameter range for our simulations \cite{Ali:2018dnd}. We have also confirmed consistency with the supersymmetric Ward identities \cite{Ali:2018fbq}.

\begin{figure}
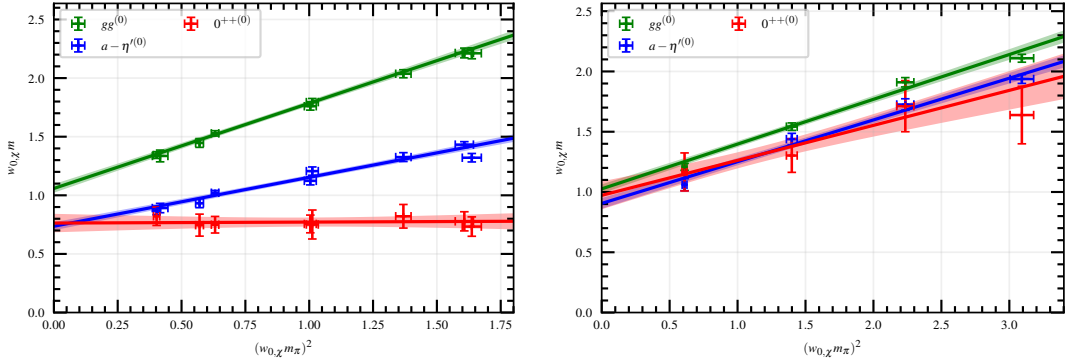

	\scalebox{0.8}{\input{figures/figureb54.pgf}}
	\scalebox{0.8}{\input{figures/figureb56.pgf}}
	\caption{Extrapolation to the chiral and continuum limit shown in the plane of fixed lattice spacing. The figure on the left hand side shows our coarsest lattice spacing $\beta=5.4$, while the figure on the right hand side corresponds to the finest considered lattice spacing $\beta=5.6$. The lightest state of the gluino-glue ($gg^{(0)}$), the scalar ($0^{++ (0)}$), and the pseudoscalar ($\aetap^{(0)}$) channel are shown as a function of the adjoint pion mass $m_\pi$ in units of the gradient flow scale $w_{0,\chi}$.\label{fig:extrchiral}}
\end{figure}
\begin{figure}
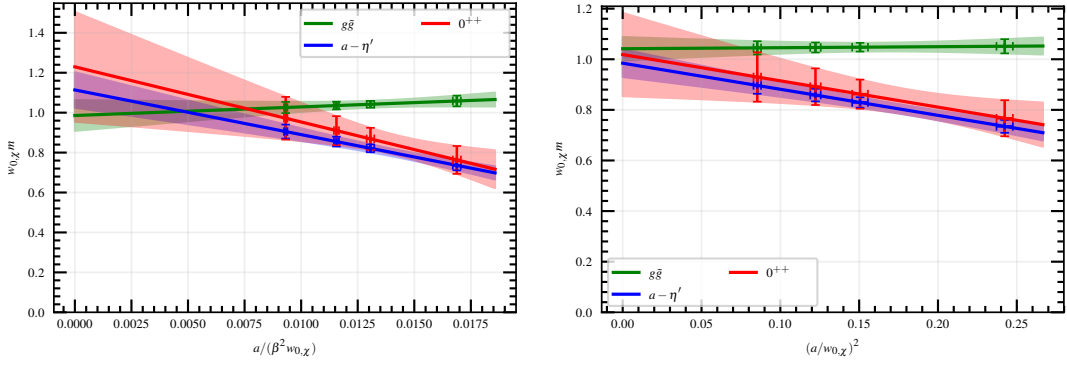

	\scalebox{.8}{\input{figures/figures_continuum_linear.pgf}}
	\scalebox{.8}{\input{figures/figures_continuum_quadr.pgf}}
	\caption{Extrapolation to the chiral and continuum limit shown in the plane of zero adjoint pion mass. The left hand side uses a linear extrapolation in the lattice spacing $a$, taking into account the one-loop improvement with the inverse coupling constant $\beta$. The right hand side is a quadratic extrapolation in $a$. All quantities are in units of the gradient flow scale $w_{0,\chi}$.\label{fig:extrcont}}
\end{figure}

We have recently completed our simulations and we have been able to extrapolate the particle spectrum to the continuum and to the chiral limit. In the previous SU(2) SYM project, we have performed the two extrapolations independently. Our most recent data for SU(3) SYM allowed to apply a more reliable approach in terms of a simultaneous two-dimensional fit of all data, see \cite{Ali:2019agk} for further details. The data and fits projected in the plane of two lattice spacings are shown in Figure~\ref{fig:extrchiral}. The supersymmetric point is reached in the chiral plane at vanishing lattice spacing (Figure~\ref{fig:extrcont}). Since we have used a one-loop improved lattice action, a functional dependence of discretization effects according to $a g^4$ is expected, where $a$ is the lattice spacing and $g$ the bare coupling constant. We have found that the our data are also in good agreement with a quadratic dependence on the lattice spacing.

The final parameter ranges are: for the pion mass $0.2<a m_{\api}<0.7$, for the lattice spacing $0.053\, \textrm{fm} < a < 0.082\, \textrm{fm}$, and for the lattice sizes from $12^3\times 24$ to $24^3\times 48$. The masses of bound states in units of the gradient flow scale $w_0$ are summarized in the following table:
\begin{center}
	\begin{tabular}{|llll|}
		\hline
		Fit  &     $w_0m_{g\tilde{g}}$ &   $w_0 m_{0^{++}}$ & $w_0 m_{\mathrm{a-}\eta'}$ \\
		\hline
		linear fit  &   0.917(91) &  1.15(30) &  1.05(10) \\
		quadratic fit &  0.991(55) &  0.97(18) &  0.950(63) \\
		% 	SU(2) SYM in~\cite{Bergner:2015adz} & 1.111(74) & 1.43(28) & 1.22(11)\\
		SU(2) SYM   & 0.93(6) & 1.3(2) & 0.98(6)  \\
		\hline
	\end{tabular}
\end{center}

To conclude, our findings for SU(3) SYM show the formation of supermultiplets of bound states and consistency with the supersymmetric Ward identities, confirming the validity of our numerical approach and providing the starting point for further investigations in two different directions. 

The first direction is a more detailed study of the properties of SYM, such as the phase transitions of this theory. At non-zero temperatures, we have found an interesting interplay between the chiral and deconfinement transitions \cite{Bergner:2014saa}. Deconfinement is absent in compactified SYM on $R^3\times S^1$, and there is a continuity towards the semiclassical regime at a small compactification radius \cite{Bergner:2014dua}. The theory at zero temperature in the chiral limit provides further interesting features that require a more detailed investigation. It is expected that spontaneous chiral symmetry breaking leads to $N_c$ different values of the gluino condensate for SU($N_c$) SYM.  

The second line of investigations is the extension of numerical studies towards more general supersymmetric gauge models, such as super-QCD. The scalar fields of these theories lead to a larger number of relevant supersymmetry breaking operators, and a more delicate tuning procedure is required.

%%%%%%%%%%%%%%%%%%%%%%%%%%%%%%%%%%%%%%%%%%%%%%%%%%%%%%%%%
\section{Exploratory studies of supersymmetric Yang-Mills theory with overlap fermions}
Ginsparg-Wilson fermions allow to define an action with an intact (modified) chiral symmetry even at non-vanishing lattice spacing \cite{GIS82}. The formulation of the Dirac operator fulfilling the Ginsparg-Wilson relation is not unique. Domain-Wall fermions provide an exact solution in the limit of an infinitely extended fifth dimension, and have been employed in some studies of the gluino condensate for gauge group SU(2) \cite{Giedt:2008xm}. Several alternative approximation schemes have been suggested in the literature. In one exploratory investigation, also the overlap operator has been considered \cite{Kim:2011fw}.

We have started to explore overlap gluinos. In our approach we implement the sign function in terms of a polynomial approximation, allowing even in the massless limit to provide a smooth and regular force for the integration of the equation of motion required by the hybrid Monte Carlo method. The simulation cost of overlap gluinos is huge, as two expensive approximations are required to compute the sign function and the square root of the determinant corresponding to the Pfaffian. In this context, a polynomial approximation avoids an additional inner inverter to the one already required to compute the rational approximation of the square root employed by RHMC, and we can reach a stable precision of the rational approximation in this way.

Several drawbacks have to be faced considering the overlap operator, mainly related to the zero modes. Effectively, each configuration with non-trivial topology has a zero mode, such that its determinant and its Monte Carlo probabilistic weight is zero. This fact would effectively induce a fixed topology in the simulations. However, zero modes are also responsible for gluino condensation, and therefore we are facing a ``zero over zero'' problem. A controlled approximation of the overlap formula can avoid zero modes as they are effectively smoothend by the polynomial approximation. 

The huge simulation cost represents a challenge for the complete investigation of the bound state spectrum of SYM with Ginsparg-Wilson fermions. However, several of our follow-up projects would benefit from preserving chiral symmetry on the lattice. Wilson fermions lead to an additive renormalization of the gluino condensate, and it is more difficult to investigate spontaneous chiral symmetry breaking and the different phases of the gluino condensate. The additive renormalization problem can be avoided using the gradient flow, but the effects of chiral symmetry breaking prevent from observing the $N_c$ phases of the gluino condensate even in the chiral limit. Therefore a detailed study of the gluino condensate requires a careful consideration of this fermion formulation.

In super-QCD and other supersymmetric gauge theories, the large number of fine-tuning parameters is reduced by the constraints of chiral symmetry. Even if fine-tuning can be handled by checking the Ward identities, the cost of a Ginsparg-Wilson implementation will be balanced by the reduced tuning costs. 

\begin{figure}
	\centering
	\includegraphics[width=0.49\textwidth]{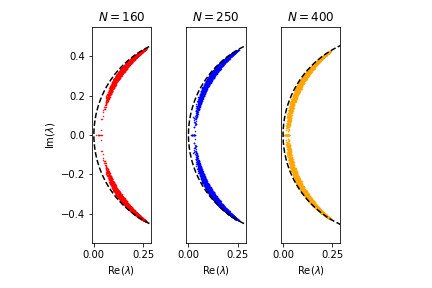}
	\includegraphics[width=0.43\textwidth]{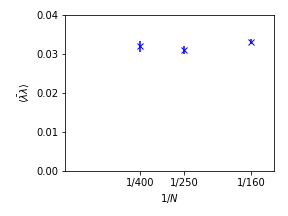}
	\caption{
	 The polynomial approximation of order $N$ of the overlap formula. Left hand side: the deviation from the exact overlap operator, whose eigenvalues lie on a circle. The eigenvalues for polynomial orders 160, 250 and 400 are shown. Right hand side: The sensitivity of the gluino condensate to the order of the approximation.\label{fig:polyoverlap}}
\end{figure} 

We have performed preliminary investigations of SU(2) SYM at $\beta=1.6$ on a small $8^4$ volume. The precision of our approximation is shown in Figure \ref{fig:polyoverlap}. The location of zeroes converges to the circle in the complex plane, corresponding to the exact overlap operator. The zero mode problem is solved by a spectral gap, which is closing for more precise approximations, as expected. Remarkably, the gluino condensate seems to be quite stable as a function of the order of the approximation, at least for the range considered in these first tests.

%%%%%%%%%%%%%%%%%%%%%%%%%%%%%%%%%%%%%%%%%%%%%%%%%%%%%%%%%
\section{Phase transitions and compactified supersymmetric Yang-Mills theory}
We have intensified our investigations  of the phase transitions in SYM, in an attempt to understand the relation between confinement and chiral symmetry breaking in this theory. We have confirmed the coincidence of the two transitions, considering also the gauge group SU(3), using the gradient flow to eliminate the difficulties with the additive renormalization of the condensate \cite{Bergner:2019dim,Bergner:2019kub}.

Another interesting aspect of the phase diagram in SYM is the absence of the deconfinement transition if periodic boundary conditions are chosen instead of the usual thermal ones. This property of SYM on $R^3\times S^1$ has led to the conjecture of continuity down to a semiclassical regime at a small radius of the compactified direction \cite{VAR}.

In our first numerical investigations of compactified SYM, we have verified the absence of the deconfinement transition \cite{Bergner:2014dua}. However, in the regime of small compactification radius, we have observed deviations from the predicted behavior. The confinement region extends towards larger masses when the compactification radius shrinks, whereas it is expected that this region gets smaller at smaller radii. 

In our most recent analysis \cite{Bergner:2018unx}
we were able to identify the difference between the observed and predicted phase boundaries as a lattice artefact stemming from the Wilson fermions. At small radius, effectively a larger number of fermion fields contribute to the dynamics driven by lattice artefacts.

\section{Conclusions}
We have obtained our final results for the low-lying bound states of SU(3) SYM using one-loop clover-improved Wilson fermions, finding evidence for restoration of supersymmetry in the continuum limit from the bound state spectrum and from the SUSY Ward identities. We have several uncertainties safely under control, like finite volume effects and the sign problem.

We have started first exploratory studies with overlap fermions based on a polynomial approximation of the sign function. This appears to be a controlled and feasible algorithm for practical simulations, which would lead to a cleaner approach for the investigations of the gluino condensate, and which reduces the fine-tuning problem for super-QCD. 

Another aspect of our investigations are the phase transitions of SYM. The coincidence of the chiral and deconfinement transition has been confirmed in our most recent simulations. The numerical results for the compactified theory support the continuity down to the semiclassical regime at small compactification radius. At a very small radius, lattice artefacts have a significant influence on the transition line. Nevertheless, at an intermediate radius, the theory should already mimic the semiclassical expectations and is an interesting candidate for further investigations.
%%%%%%%%%%%%%%%%%%%%%%%%%%%%%%%%%%%%%%%%%%%%%%%%%%%%%%%%%%%%%%%%%%%%%%%%
\section*{Acknowledgements}\vspace*{-0.3cm}
We thank M.~\"Unsal for his contributions to the investigations of compactified SYM.
The authors gratefully acknowledge the Gauss Centre for Supercomputing
e.\,V.\, (www.gauss-centre.eu) for funding this project by providing
computing time on the GCS Supercomputers JUQUEEN, JURECA, and JUWELS at J\"ulich Supercomputing
Centre (JSC) and SuperMUC at Leibniz Supercomputing Centre (LRZ). Further
computing time has been provided the compute cluster PALMA of the University of M\"unster. This work is
supported by the Deutsche Forschungsgemeinschaft (DFG) through the Research
Training Group ``GRK 2149: Strong and Weak Interactions - from Hadrons to
Dark Matter''. G.~Bergner acknowledges support from the Deutsche
Forschungsgemeinschaft (DFG), Grant No.\ BE 5942/2-1. S.~Ali acknowledges
financial support from the Deutsche Akademische Austauschdienst (DAAD).

\end{document}